\documentclass[]{spie}  
\usepackage[]{graphicx}

\usepackage{aasmacrospk}
\DeclareGraphicsExtensions{.pdf,.png,.jpg,.eps}

\def\simgt{\ {\raise-.5ex\hbox{$\buildrel>\over\sim$}}\ }
\def\simlt{\ {\raise-.5ex\hbox{$\buildrel<\over\sim$}}\ }

\def \h70{{h_{70}}}

\title{Advanced Technologies and Instrumentation and the National Science Foundation} 

\author{Peter Kurczynski\supit{a,b} and James E. Neff\supit{a}
\skiplinehalf
\supit{a}National Science Foundation, 2415 Eisenhower Ave, Alexandria , Virginia, USA; \\
\supit{b}Rutgers, the State University of New Jersey, Piscataway, New Jersey, USA
}

\authorinfo{Send correspondence to P.K. at pkurczyn@nsf.gov, Telephone 703-292-7248.  The views expressed here are solely those of the authors and do not reflect the views of the National Science Foundation.}

\begin{document}

\maketitle 

\begin{abstract}
Over its more than thirty-year history, the Advanced Technologies and Instrumentation (ATI) program has provided grants to support technology development for ground-based astronomy.  Research from this program has advanced adaptive optics, high resolution and multi-object spectroscopy, optical interferometry and synoptic surveys, to name just a few.  Previous and ongoing scientific advances span the entire field of astronomy, from studies of the Sun to the distant universe.  Through a combination of literature assessment and individual case studies, we present a survey of ATI funded research for optical-infrared astronomy.  We find that technology development unfolds over a time period that is longer than an individual grant.  A longitudinal perspective shows that substantial scientific gains have resulted from investments in technology.
\end{abstract}



\section{INTRODUCTION}
\label{IntroductionSection}

Ever since the dawn of modern astronomy, our understanding of the universe has advanced through increasingly accurate and precise observation.  Naked-eye observations of the planets by Tycho Brahe and Johannes Kepler circa 1609, which were accurate to about 0.1 degrees, proved that the Earth revolves around the Sun and that planetary orbits are elliptical.\footnote{The heliocentric theory of Copernicus, published earlier in 1543, used circular orbits to explain planetary positions on the sky and was in error by about the same $\sim$5 degrees as the contemporaneous model of deferents and epicycles to Earth-centered, planetary orbits [\citenum{Kuhn1957}].}  Today's Laser Interferometer Gravitational-Wave Observatory (LIGO) measures the movement of test masses to less than an atomic diameter; this extraordinary precision recently enabled the discovery of gravitational waves from colliding black holes [\citenum{PhysRevLett.116.061102}] and neutron stars [\citenum{2017PhRvL.119p1101A}] and gave birth to a new field of gravitational wave astronomy.  

Our ability to accurately and precisely observe the universe has evolved in lock-step with the development of technology.  Today's development and application of specialized technologies for astronomy is supported in part by federal funding agencies.  The National Science Foundation supports the development of new technologies within the Division of Astronomical Sciences through the Advanced Technologies and Instrumentation (ATI) program, as well as through other programs.
This paper presents an historical overview of the ATI program with the goal of addressing the impact of technology development supported by ATI over the years.

The effects of a particular research program can be assessed through analysis of the scientific literature.  Direct effects can be attributed to research publications that follow directly from an award, which investigators are required to acknowledge in their publications.  However, additional indirect effects may be more far-reaching, though harder to quantify.  A new technological solution to a particular problem may enable a wide range of scientific investigations, spur additional innovation in the field or comprise a critical role in a larger initiative.  These impacts may not be fully apparent until years or even decades after an initial award.   Also important are the education and training opportunities that a research program provides.  The tools and methods of astronomy are increasingly specialized; opportunities to gain mastery of these methods are rare and precious.  Such indirect impacts of awards will be illustrated through particular examples over the history of the program.  Section 2 describes the ATI program and places it in context of other NSF programs.  Section 3 presents the results of automated literature citation analysis and comparison to a ``pure science" program that does not emphasize technology development.  Section 4 presents a narrative history of the program and some of the illustrative awards and their resulting impact.  The information here is obtained from the published literature as well as from conversations with individual investigators and experts in the field.  Section 5 summarizes the results and presents conclusions.  Although the ATI program actively supports projects aimed at diverse wavebands from radio-waves through optical bandpasses, for the sake of brevity and topical relevance to the conference session, this paper will tend to focus on awards for optical-IR waveband investigations.

\section{Advanced Technologies and Instrumentation}
\label{ATISection}
The National Science Foundation (NSF) was established in 1950 as a federal agency for fostering civilian scientific research [\citenum{Mazuzan1992}]. NSF is funded through Congressional appropriations that amounted to approximately \$7.5B in Fiscal Year (FY) 2017; it has a portfolio that spans the entire range of research and education in science and engineering.  The Division of Astronomical Sciences has an annual budget of approximately \$240M (FY2017) that supports ground-based astronomical facilities as well as individual investigator grants.  
   \begin{figure}
   \begin{center}
   \begin{tabular}{c}
   \includegraphics[height=7cm]{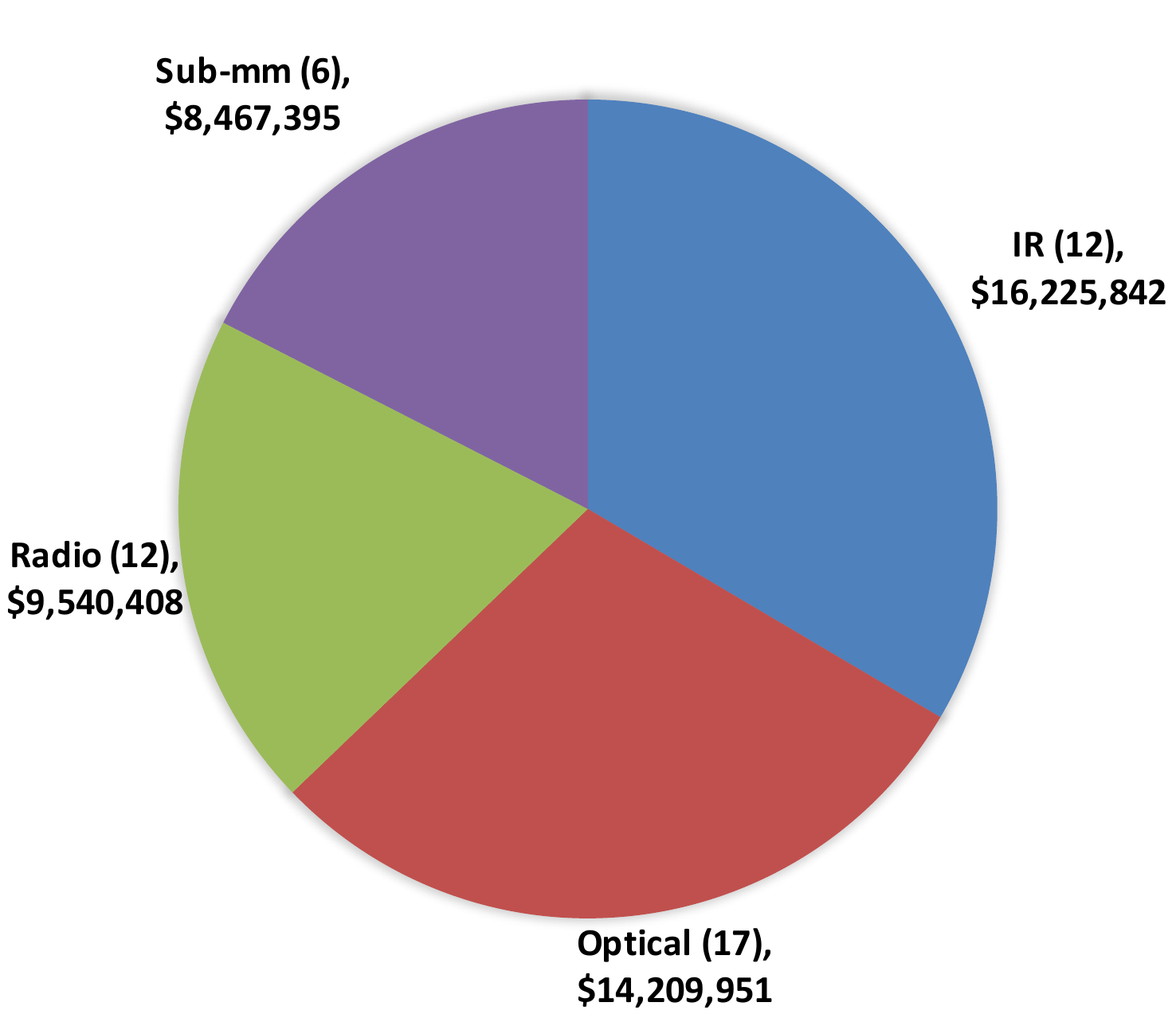}
   \includegraphics[height=7cm]{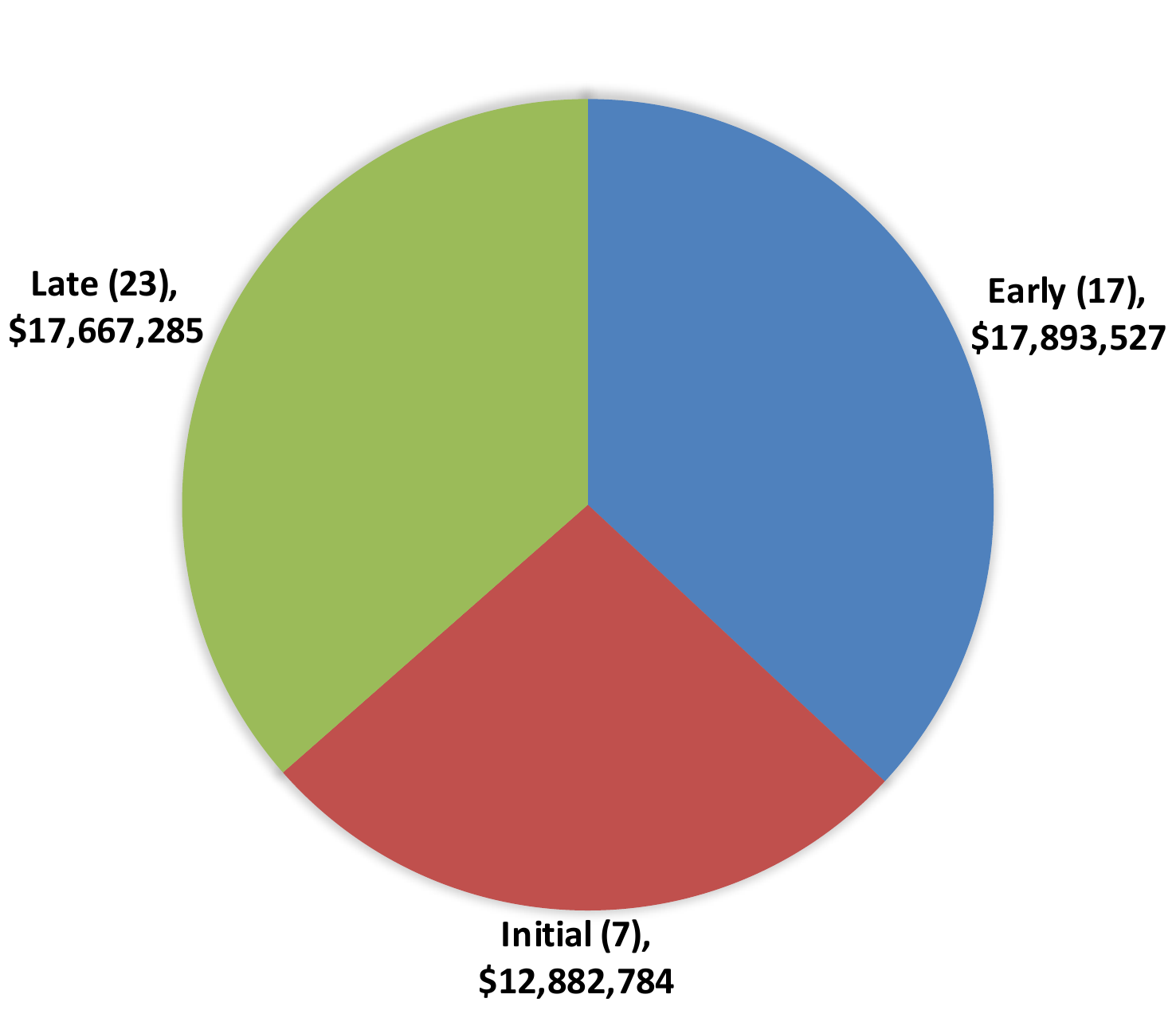}
   \end{tabular}
   \end{center}
   \caption[ActiveAwardsByWavebandAndTechDevp] 
   { \label{fig: ActiveAwardsByWavebandAndTechDevp} 
Summary of active ATI awards by waveband {\bf (left panel)} and technology development phase {\bf (right panel)}.  Size of each pie segment indicates the amount awarded to date (also shown in dollars); the corresponding number of awards are indicated in parentheses.  Awards are distributed across radio, sub-millimeter, IR and optical wavebands.  The active portfolio is broadly diversified among initial concepts and more mature technologies, as discussed in the text.}
   \end{figure} 

The bulk of astronomical research that is supported by individual investigator programs are included within the Astronomy and Astrophysics Grants (AAG) program. Grants are also awarded to support graduate student and post doctoral fellowships, undergraduate training, prestigious CAREER and other awards.\footnote{For a complete listing, see  {\tt https://www.nsf.gov/funding/programs.jsp?org=AST}}.  The Mid-Scale Innovations Program (MSIP) and the smaller ATI program each support new instrumentation and technology development for astronomy.  Deployment of existing technology for astronomical applications is supported through the Foundation-wide Major Research and Instrumentation (MRI) program.  The ATI and MSIP programs focus on technology development and/or instrumentation in support of specific scientific objectives.  ATI has supported research into adaptive optics, high resolution and multi-object spectroscopy, optical interferometry and synoptic surveys, to name just a few.   Basic information on all NSF awards are publicly available.\footnote{see {\tt https://www.nsf.gov/awardsearch/}} Such information includes the NSF award ID (a seven digit number beginning with the two-digit year of the award; e.g. award 8911701 was awarded in FY 1989), principal investigator name and organization (institutional affiliation), title and abstract, program name, start and end dates and awarded amount to date.  These  public data for the ATI program from 1987 - 2016 form the basis for the analyses presented here.  Hereafter, where specific awards are mentioned, they will be referred by their NSF award ID followed by the principal investigator last name, e.g. 8911701/Angel. 
   \begin{figure}
   \begin{center}
   \begin{tabular}{c}
   \includegraphics[height=7cm]{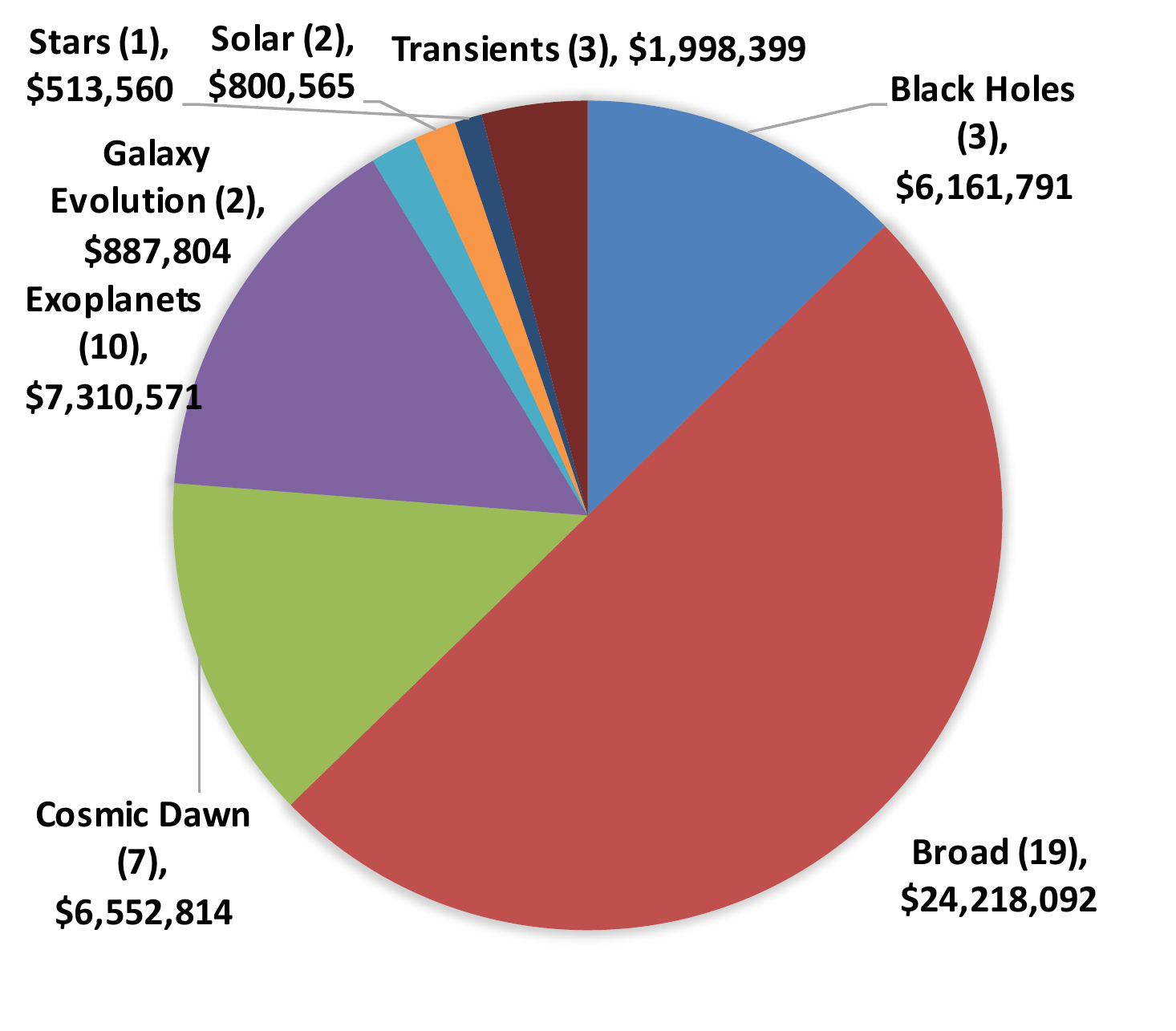}
   \includegraphics[height=7cm]{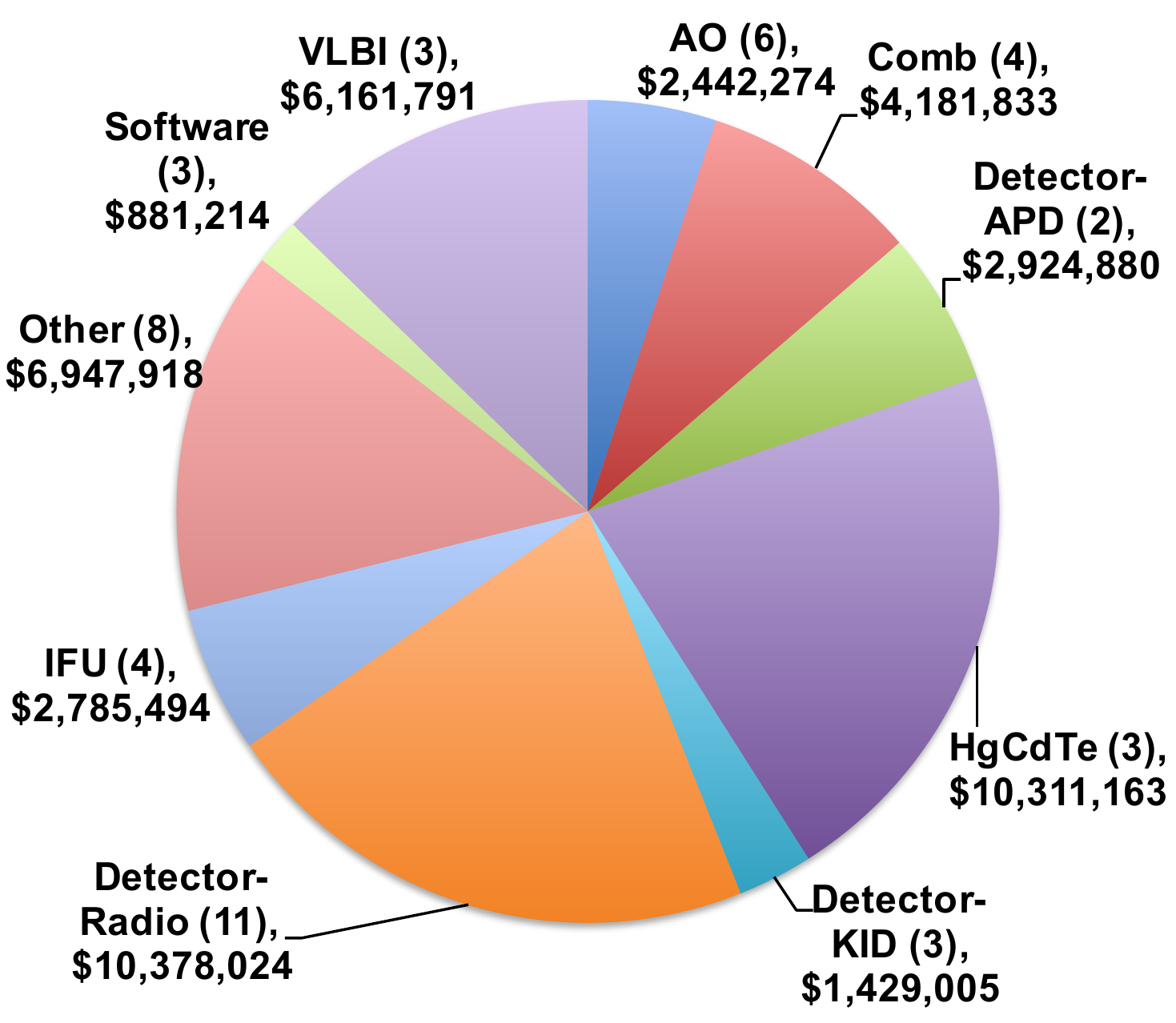}
   \end{tabular}
   \end{center}
   \caption[ActiveAwardsByScienceAndTechnology] 
   { \label{fig: ActiveAwardsByScienceAndTechnology} 
Summary of active ATI awards by science category {\bf (left panel)} and technology type {\bf (right panel)}.  Same format as Figure \ref{fig: ActiveAwardsByWavebandAndTechDevp}.  Awards are distributed across a range of science topics, with the largest segment of awards supporting multiple/broad science goals.  A variety of technologies are supported including Very Long Baseline Interferometry (VLBI), Adaptive Optics (AO), laser frequency combs (Comb), Avalanche Photo-Diode detectors (APDs), HgCdTe IR detectors, Kinetic Inductance Detectors (KIDs), Radio waveband technologies (Radio), Integral Field Units (IFUs), other un-categorized technologies (Other) as well as software.}
   \end{figure} 

A summary of the active ATI portfolio is illustrated in Figures \ref{fig: ActiveAwardsByWavebandAndTechDevp} and \ref{fig: ActiveAwardsByScienceAndTechnology}.  These data were downloaded on October 1, 2016 and therefore do not include awards made in FY 2017.  Active awards were categorized by waveband, science objective, technology type and technology development phase based on information in the title and abstract.  Technology development phase is intended to classify the maturity of a technology and it is analogous to Technology Readiness Levels used by NASA [\citenum{2014SPIE.9143E..13P,Perez2016}].  Awards were classified as ``initial" phase if the proposed research was an initial concept development that would demonstrate a process or technology but would not lead to a full engineering prototype at the end of the award period.  Awards were classified as ``early" phase if they planned to deliver an engineering prototype to demonstrate a technology but would not deliver new science data at the end of the award period.  ``Late" technology development awards would produce new science data at the end of the award period (e.g. by installing an instrument at a telescope and using it for science).  

As illustrated in Figure \ref{fig: ActiveAwardsByWavebandAndTechDevp}, the active ATI portfolio is broadly diversified among wavebands and technology development phase; about two-thirds of the program consists of projects to study optical-IR wavebands.  Figure \ref{fig: ActiveAwardsByScienceAndTechnology} shows that ATI covers a broad range of science and technology areas.  Exoplanets and cosmic dawn (cosmic dark ages, epoch of reionization and early structure formation) are notable science categories.  Three awards to study black holes (including one award that is cross-listed as an MRI award) are actually in support of one over-arching research program - the Event Horizon Telescope.  Technology categories are similarly broad; the categories most relevant for optical-IR astronomy include adaptive optics (AO; high spatial resolution imaging), laser frequency combs (an enabling technology for high precision radial velocity measurements and exoplanet research), IR detectors (including avalanche photo-diodes, APDs, Mercury Cadmium Telluride (HgCdTe), and Kinetic Inductance Detectors, KIDs) and integral field units (IFUs; spatially resolved spectrographs).

\section{LITERATURE AND CITATION ANALYSIS}
\label{LiteratureSection}
An automated search of the astronomical literature was done to asses the direct impact of ATI awards.  Because investigators are required to acknowledge their grants in publications that follow from supported research, it is possible to connect specific research products to particular awards using publicly available literature repositories.  Publicly available NSF records of ATI include 496 awards from 1987 - 2016.  Each award was searched in the astronomical literature using the Astrophysics Data System (ADS).  Code was written in Python to retrieve records from the database for use as input ADS search criteria.  A publicly available interface was used to interact with ADS.\footnote{available at {\tt  https://github.com/andycasey/ads}}  For each award record, ADS was searched for papers with the following criteria:  (1) the award ID in the full text of the paper, (2) the Principal Investigator (PI) name as an author (3) publication date range specified by the award effective date and 2017.  The search was restricted to peer-reviewed publications.  The decision to exclude conference presentations, even though many instrument papers are published in non-refereed proceedings, was chosen to make the results of the automated search conservative, but more robust.  The number of acknowledging papers and the three most highly cited papers that acknowledge the award as well as their number of citations were saved from each search.  These search results were stored in the PostgreSQL database for analysis. 
%
\begin{table}[h]
\caption{Distribution of acknowledgements of ATI and PLA awards in peer-reviewed literature since 1987.  Column (1) indicates the number of acknowledgements found by the automated routine.  Columns (2, 3) indicate the number of awards that have at least as many acknowledgements as indicated in Column (1)  in the ATI and PLA programs respectively.  The bottom row indicates total number of awards in each program} 
\label{tab:AcknowledgementTable}
\begin{center}       
\begin{tabular}{|c|c|c|} 
\hline
\rule[-1ex]{0pt}{3.5ex}  {\bf Acknowledgments} & {\bf ATI} & {\bf PLA} \\
\rule[-1ex]{0pt}{3.5ex}  {\bf (1)} & {\bf (2)} & {\bf (3)} \\
\hline
\rule[-1ex]{0pt}{3.5ex}  20 & 3 & 3   \\
\hline
\rule[-1ex]{0pt}{3.5ex}  10 & 18 & 16  \\
\hline
\rule[-1ex]{0pt}{3.5ex}  5 & 51 & 35  \\
\hline
\rule[-1ex]{0pt}{3.5ex}  2 & 140 & 83  \\
\hline
\rule[-1ex]{0pt}{3.5ex}  1 & 216 & 138  \\
\hline
\rule[-1ex]{0pt}{3.5ex}  Total & 496 & 445  \\
\hline 
\end{tabular}
\end{center}
\end{table} 

For comparison, an identical search was performed on awards from the Planetary Astronomy (PLA) program, which was recently renamed Solar and Planetary Grants.  PLA was selected as a control sample because of its comparable budget, number of awards and availability of historical data over a similar range of dates as ATI.  This sample included 445 awards between 1989 - 2016.  Planetary astronomy funds pure science proposals without an emphasis on instrumentation; thus a comparison with ATI may address whether there is a difference in impact for astronomical technology development as opposed to pure science.

Table \ref{tab:AcknowledgementTable} summarizes the results of the automated literature search.  It tabulates the number of awards that were acknowledged and how many times they were acknowledged in the peer-reviewed literature.  These results show that 44\% (216/496) of ATI awards are acknowledged at least once in the peer-reviewed literature, compared with 31\% (138/445) in PLA.  Multiple acknowledgements were found with less frequency, and a handful of awards from each program were acknowledged more than twenty times.   Table \ref{tab:AcknowledgementTable} demonstrates that ATI and PLA awards are acknowledged with approximately the same frequency in the peer-reviewed astronomical literature despite the fact that ATI awards emphasize technology development, whereas PLA awards are for ``pure science." 

Table \ref{tab:WidelyAcknowledgedAwardsAndCitedPapers} illustrates the most widely acknowledged awards and the most highly cited paper among those acknowledging awards.  The most widely acknowledged ATI award, 0906060/Baranec supported deployment of a micro-machined deformable mirror for the adaptive optics system on the Robo-AO robotic observatory.  This award was acknowledged 31x in the peer-reviewed literature.  The most highly cited paper from this award was cited 79x [\citenum{2014ApJ...790L...8B}].  The other two most widely acknowledged awards, 0705139/Ge and 1006676/Mahadevan supported technology for studying exoplanets.  These ATI awards have been broadly aligned with the objectives of the most recent astronomy \& astrophysics decadal review, for which ``seeking nearby, habitable planets" was one of three main science objectives [\citenum{NWNH2010}].
%
%
\begin{table}[h]
\caption{Three most widely acknowledged ATI awards, and two other awards that were less acknowledged but were nevertheless impactful.  Column (1) is the NSF award ID.  Column (2) is the Principal Investigator (PI) of the award.  Column (3) is the number of acknowledgements found in the automated search.  Column (4) is the ADS bib code for the most highly cited paper among the acknowledging publications for this award.  Column (5) is the number of citations that the most cited publication has in ADS} 
\label{tab:WidelyAcknowledgedAwardsAndCitedPapers}
\begin{center}       
\begin{tabular}{|c|c|c|c|c|} 
\hline
\rule[-1ex]{0pt}{3.5ex}  {\bf Award ID} & {\bf PI} & {\bf Acknowl.}  &{\bf Most Cited} &{\bf Citations} \\
\rule[-1ex]{0pt}{3.5ex}  {\bf (1)} & {\bf (2)} & {\bf (3)}  &{\bf (4)} &{\bf (5)} \\
\hline
\rule[-1ex]{0pt}{3.5ex} 0906060	&	Baranec, Christoph	&	31	&	2014ApJ...791...35L	&		79\\
\rule[-1ex]{0pt}{3.5ex} 0705139	&	Ge, Jian			&	27	&	2011ApJ...728...32L	&		29\\
\rule[-1ex]{0pt}{3.5ex} 1006676	&	Mahadevan, Suvrath	&	22	&	2014Sci...345..440R	&		68\\
\hline
\rule[-1ex]{0pt}{3.5ex} 9731180	&	Elston,	Richard		&	5	&	2003AJ....125.2029M	&		112\\
\rule[-1ex]{0pt}{3.5ex} 0441069	&	Tyson, J. Anthony	&	1	&	2014JInst...9C7010T	&		0\\
\hline
\end{tabular}
\end{center}
\end{table} 

Table \ref{tab:CitationsAndImpact} shows the five most highly cited papers from the last thirty years of the ATI and PLA programs.  The most highly cited ATI paper presents first results of the Degree Angular Scale Interferometer (DASI) measurement of the cosmic microwave background angular power spectrum [\citenum{2002ApJ...568...38H}].  The receivers, the dewars, the local oscillator distribution scheme, and the correlator and were all made with support from  9413935/Readhead, which also supported development and construction of a sister instrument, the Cosmic Background Imager.  

%
%
\begin{table}[h]
\caption{The most highly cited papers that acknowledge ATI and PLA awards.  Columns (1,2) indicate the NSF award ID and PI last name respectively.  Column (3) is the ADS identifier for the paper.  Column (4) is the year of publication of the paper.  Column (5) is the number of citations.  Column (6) is the median number of citations for the year of publication.  Column (7) is the impact, defined as citations / median.  * indicates an unreliable value as discussed in the text.} 
\label{tab:CitationsAndImpact}
\begin{center}       
\begin{tabular}{|c|c|c|c|c|c|c|} 
\hline
\rule[-1ex]{0pt}{3.5ex}	{\bf Award}	&	{\bf PI}	&	{\bf Bibcode}	&	{\bf Year}	&	{\bf Citations}	&	{\bf Median}	&	{\bf Impact}	\\
\rule[-1ex]{0pt}{3.5ex}	{\bf (1)}	&	{\bf (2)}	&	{\bf (3)}	&	{\bf (4)}	&	{\bf (5)}	&	{\bf (6)}	&	{\bf (7)}	\\
\rule[-1ex]{0pt}{3.5ex}		&		&	{\bf ATI}	&		&		&		&		\\
\rule[-1ex]{0pt}{3.5ex}	9413935	&	Readhead	&	2002ApJ...568...38H	&	2002	&	769	&	36	&	21.4	\\
\rule[-1ex]{0pt}{3.5ex}	0096913	&	Carlstrom	&	2002ARA\&A..40..643C	&	2002	&	561	&	36	&	15.6	\\
\rule[-1ex]{0pt}{3.5ex}	8822465	&	McCarthy	&	1991ApJS...77..417K	&	1991	&	495	&	25	&	19.8	\\
\rule[-1ex]{0pt}{3.5ex}	0904607	&	Townsend	&	2013ApJS..208....4P	&	2013	&	489	&	9*	&	54.3*	\\
\rule[-1ex]{0pt}{3.5ex}	9203336	&	McCarthy	&	1993AJ....106..773H	&	1993	&	420	&	27	&	15.6	\\
\rule[-1ex]{0pt}{3.5ex}	 	&	 	&	 	&	 	&	 	&	 	&	 	\\
\hline
\rule[-1ex]{0pt}{3.5ex}		&		&	{\bf PLA}	&		&		&		&		\\
\rule[-1ex]{0pt}{3.5ex}	9120599	&	Begelman	&	1994ApJ...421..153S	&	1994	&	860	&	27	&	31.9	\\
\rule[-1ex]{0pt}{3.5ex}	8857365	&	Wisdom	&	1991AJ....102.1528W	&	1991	&	675	&	25	&	27.0	\\
\rule[-1ex]{0pt}{3.5ex}	9530590	&	Heiles	&	2003ApJ...586.1067H	&	2003	&	341	&	28	&	12.2	\\
\rule[-1ex]{0pt}{3.5ex}	9973057	&	Tedesco	&	2002AJ....123.1056T	&	2002	&	310	&	36	&	8.6	\\
\rule[-1ex]{0pt}{3.5ex}	9714275	&	Lin	&	2001ApJ...548..466B	&	2001	&	249	&	31	&	8.0	\\
\rule[-1ex]{0pt}{3.5ex}	 	&	 	&	 	&	 	&	 	&	 	&	 	\\
\hline
\end{tabular}
\end{center}
\end{table} 

In addition to the number of citations and the year of publication for each paper, Table \ref{tab:CitationsAndImpact} also shows the median number of citations for that year of a paper published in {\it The Astronomical Journal}.  This median value was obtained from a search of ADS for all papers in {\it The Astronomical Journal} for the year in question.  Notably, median citations in this journal are in the 20-30 range throughout the 1990's and 2000's, but they decline significantly after 2009.  Therefore, after 2009 median citations in this journal should not be taken as representative of the entire field (applies to one paper in the table, entry marked with an asterisk).  

An impact factor is defined as the number of citations divided by the median number of citations for that year, and it is tabulated in the last column of Table \ref{tab:CitationsAndImpact}.  The tabulated impact factor after 2009 is considered unreliable due to uncertainty in the median.  The impact factor normalizes for an age effect, whereby older papers have more citations simply because they are older.  The distribution of such normalized citation counts has been found to follow the same log-normal distribution across a wide range of scientific disciplines [\citenum{Fortunato:2018ih,2008PNAS..10517268R}].  The papers shown here are in the high-impact tail of this distribution.  

The admittedly small sample in Table \ref{tab:CitationsAndImpact} suggests that ATI and PLA papers have comparable impact.   Excluding the ATI paper published after 2009, these sub-samples of the most highly cited papers from each program have the same average impact factor ($\approx$18), although the two highest impact papers are found in the PLA program.

Several factors may affect the accuracy of literature acknowledgements and resulting paper citations.  A false acknowledgement (``false positive" or type I error) would be an instance of the automated search routine reporting an acknowledgement in a paper when in fact no such acknowledgement exists.  No such errors were found among manual searches of several dozen papers.  A ``false negative" (type II error) occurs when a paper that was derived from grant-supported research is not found by the automated acknowledgement search algorithm.  There are several possible explanations for such a false negative:  A non-ads acknowledgement is defined as a product of research that is published outside of the accessible ADS search; these publications may be books or journals that are not included within ADS.   A delinquent acknowledgement is defined as a missing or incomplete acknowledgement of an NSF award.  Due to such false negatives, the automated search should be multiplied by approximately a factor of two to reflect the actual publication rate from these awards. 

\section{DISCUSSION:  ATI THROUGH THE YEARS}
\label{DiscussionSection}

Inspection of the titles and abstracts of ATI awards reveals broad trends over time.  ATI awards in the 1980's and 1990's were notable for disseminating Charge Coupled Device (CCD) cameras to astronomical observatories for research and teaching.  Although they were not initially conceived for astronomy, CCDs were applied to astronomy during this time period and their deployment to observatories was transformative; they are discussed extensively elsewhere, e.g. [\citenum{McClean2010}].  Mauna Kea, Lick and Lowell as well as many smaller observatories received CCD cameras through ATI awards in this time period.  

Adaptive Optics (AO) have been developed and disseminated throughout the history of the ATI program.  AO was conceived in 1953 as a means of correcting images for the time-varying distortions caused by atmospheric turbulence  [\citenum{1953PASP...65..229B}]; it was subsequently developed for military and civilian applications [\citenum{1993ARA&A..31...13B}].   

ATI supported the development of laser guide stars and the deployment of AO instrumentation at Mauna Kea, MMT, Mt.~Wilson and Apache Point observatories beginning in the 1990's.  Edward Kibblewhite received four ATI awards between 1993 and 2008 (9256606, 9421406, 9731169, 0837646); this program developed the sum frequency laser guide star (with Lincoln Labs/Tom Jayes).  Although not adopted in the current (Topica) lasers, this program was a key evolution to the current lasers (M. Chun, Personal Communication, October 5, 2017).  

ATI award 9319004/Roddier supported development of an AO system on the Canada France Hawaii Telescope, which was then used to study circumstellar environments and protoplanetary disks.  This research program pushed both AO system development and the science observations that use AO.  ATI award 0906060/Baranec funded the first robotic laser adaptive optics system (Robo-AO) [\citenum{2012SPIE.8447E..04B}].  Robo-AO has been used for 37 refereed publications by 27 unique first authors; results of these studies include AO imaging of nearly all of the Kepler candidate exoplanet hosts and all known stars within 25 pc (C. Baranec, Personal Communication, October 4, 2017).  

The AO system on the W.M. Keck Observatory telescope has been used extensively; highlights include imaging volcanic eruptions on Io, e.g.~[\citenum{2005Icar..176...96M}], astrometric monitoring of the orbits of stars near the galactic center that confirm the existence of a supermassive black hole [\citenum{2008ApJ...689.1044G}] and studies of proto-planetary formation and exoplanets; while not necessarily following directly from ATI awards, these and other AO studies benefitted indirectly from ATI-funded efforts.  

Multi-Conjugate Adaptive Optics (MCAO) uses multiple wavefront sensors to achieve wide-field correction; notably 1407957/Goode recently led to the first demonstration of  MCAO in solar imaging [\citenum{2017A&A...597L...8S}].  Extreme Adaptive Optics (ExAO) is a recent evolution of single-conjugated AO systems to achieve high contrast at short separations from a bright star for exoplanet searches [\citenum{2016ASSL..439...17M}].  The MagAO system on the Magellan/Clay telescope was supported by ATI (1206422/Close), and its upgrade to an ExAO system is supported by an active ATI award (1506818/Males).  Similarly an active ATI award is supporting the first AO-fed integral-field spectrograph for the thermal IR on the twin 8.4 m Large Binocular Telescope (1608809/Hinz).

ATI has also supported high resolution spectroscopy, particularly for exoplanet studies.  High precision radial velocity measurements of stars provides a key method for studying exoplanets [\citenum{2016PASP..128f6001F}].  This approach was first pioneered from an ATI award (8919634/Marcy), which used an iodine reference cell for precise wavelength calibration of the spectrograph [\citenum{1992PASP..104..270M}].  This method can be used without a special purpose spectrograph and it was adopted by others, e.g. the HIRES instrument at the W. M. Keck Observatory [\citenum{1994SPIE.2198..362V}].

A competing technology for high precision radial velocity measurements, the cross correlation method, uses a special purpose fiber-fed spectrograph.  Because of the fiber, the spectrograph can be isolated and temperature controlled, and the resulting stability has proved to be critical.  The European-funded ELODIE spectrograph at Observatoire de Haute-Provence, France used this approach [\citenum{1996A&AS..119..373B}]; it made the first detection of a Jupiter-like exoplanet around a solar-type star (51 Pegasi) [\citenum{1995Natur.378..355M}].  The first dedicated radial velocity spectrograph was the European-funded High Accuracy Radial Velocity Planet Searcher (HARPS) on the European Southern Observatory (ESO) 3.6 m telescope at La Silla [\citenum{2003Msngr.114...20M,2002Msngr.110....9P}].  This instrument demonstrated the importance of vacuum-enclosed, temperature-stabilized operation.  It has been one of the leading instruments in this field since 2005.  

Wavelength calibration in the fiber-fed spectrographs was initially accomplished via simultaneous reference from a Thorium-Argon lamp; however, in the past decade laser frequency combs have emerged as a critical new technology.  Laser frequency combs use femtosecond mode-locked lasers to produce a repetitive train of narrow emission lines, and are in principle a nearly ideal wavelength calibrator [\citenum{2007MNRAS.380..839M}].  The 2005 Nobel Prize in physics was awarded for pioneering work in this field.  The first application to astronomical measurements was accomplished by a European group [\citenum{Steinmetz2008}]; however, ATI-funded researchers were among the early adopters of this technology (e.g. 0905214/Walsworth; 0906034 and 1310875/Osterman).  Although laser frequency comb technology has matured to the point of being commercially available [\citenum{2014SPIE.9147E..1CP}], it is still an area of active research with promising new approaches to comb generation (S. Diddams, Personal Communication, October 12, 2017).  The current state of the art allows precision radial velocity measurements of $\sim$0.8 -- 1~m~s$^{-1}$; however, the signal from Earth-like habitable zone exoplanets is expected to be closer to $\sim$10 cm s$^{-1}$ [\citenum{2016PASP..128f6001F}].  Technological advances are needed to close this gap, and ATI currently funds active efforts in this field.

Detector development has been extensively supported by the ATI program.  An example of optical detector development includes the Large Synoptic Survey Telescope, (LSST), for which focal plane development was initiated with an ATI award (0243144/Tyson).  This award was the first NSF investment in what ultimately became the highest-priority for ground-based astronomy, according to the last decadal review.  The ATI award came at a critical phase in the growth of the project; without this essential seed funding, the focal plane may not have ever been developed (J.~A.~Tyson, Personal Communication, September 24, 2017).  As a result, although acknowledged sparsely in the literature (see Table \ref{tab:WidelyAcknowledgedAwardsAndCitedPapers}), this award emerges as broadly impactful.

IR detectors were primarily developed for military and commercial applications; an historical perspective of their use in astronomy is provided in [\citenum{2007ARA&A..45...43L}].  The most successful detectors to date have been based on mercury cadmium telluride (HgCdTe) semiconductor, for which development began in the 1980's.  Teledyne Imaging Sensors (formerly Rockwell Scientific Company) developed processes of HgCdTe growth by molecular beam epitaxy on cadmium zinc telluride (CdZnTe) substrates [\citenum{Zandian2003}].  Independently, Raytheon Vision Systems developed processes of HgCdTe growth on silicon and CdZnTe substrates [\citenum{Reddy2011}]; potential advantages of the latter approach include substantially reduced cost by using established silicon processing technology.

Most of IR astronomy would not exist without NASA and NSF support.  NASA substantially funded HgCdTe detector development;  such detectors were used in the NICMOS and WFC3 instruments aboard the {\it Hubble Space Telescope}.  Later generation HAWAII 2RG detectors are installed in the {\it James Webb Space Telescope} as well as on ground-based instruments.  There may be 30-50 HAWAII 2RG detectors deployed at ground-based observatories worldwide (D. Hall, Personal Communication, December 1, 2017); it is unclear how many of these deployments were funded by ATI awards.  

ATI awardees partnered with the industrial collaborators to characterize devices and tailor them to the needs of astronomy.  The largest single ATI award supported development of the HAWAII 4RG mosaic camera (0804651/Hall; \$7M, made with American Recovery \& Reinvestment Act funds) in collaboration with Teledyne.   Results of 0804651/Hall have been reported in conference proceedings [\citenum{2016SPIE.9915E..0FZ,2016SPIE.9915E..0WH}] (excluded from the automated ADS search mentioned above).  Similarly, ATI awards 1207827 and 1509716/Figer constituted substantial investments in the competing Raytheon technology.  At present the Raytheon detectors have difficulty reaching sufficiently low dark current [\citenum{Hanold2012}].  However, if further development can meet the dark current specification, Raytheon detectors could be transformative.

Traditional IR detectors are charge-integrating devices, and they are constrained by read-noise in fast readout applications.  As an alternative, avalanche photodiode arrays have been developed with support from ATI (1106391/Hall) and other funding sources in collaboration with industrial partner Leonardo (formerly Selex ES).  The Selex Advanced Photodiode HgCdTe Infrared Array (SAPHIRA) has emerged as a leading detector for near-infrared wavefront sensing in AO [\citenum{Hall2016b,Atkinson2018}].  Low read-noise, fast time response detectors will be crucial for pushing the limits in exoplanet studies [\citenum{2015PASP..127..890J}].

Another promising technology for low read-noise, fast time response detectors are Microwave Kinetic Inductance Detectors (MKIDs) [\citenum{2003Natur.425..817D}].  Unlike CCDs or photodiodes, these detectors rely upon thin superconducting films that are arranged in a microwave resonant circuit.  Incident photons break Cooper pairs and generate quasiparticles which change the complex impedance and hence the resonance of the circuit.  High quality factor resonant circuits (operating at cryogenic temperatures) can be sensitive to individual photons.  Individual detector elements can be made with unique microwave resonance frequencies and multiplexed into large arrays with simple readouts.  This technology has broad application in addition to optical-IR astronomical detectors [\citenum{2012OExpr..20.1503M}].  

The first application of an MKID detector to ground-based, optical-IR astronomy was the Array Camera for Optical to Near-IR Spectrophotometry (ARCONS) [\citenum{2013PASP..125.1348M,2015ApJS..219...14V}].    Its successor, the DARK-speckle Near-infrared Energy-resolving Superconducting Spectrophotometer (DARKNESS) was supported by an ATI award (1308556/Mazin); it was recently installed behind the PALM-3000 adaptive optics system and the Stellar Double Coronagraph at Palomar Observatory [\citenum{2018PASP..130f5001M}].  The sensitivity and fast time response of MKIDs makes them amenable to ExAO and speckle imaging/nulling applications, which are the subjects of active ATI supported investigations.

Multi-object spectroscopy has been developed through ATI awards.  The FLoridA Mulit-object Infrared Grism Observational Spectrometer (FLAMINGOS) system was the first cryogenic, multi-object spectrograph, and it was developed through an ATI award (9731180/Elston).  This instrument accomplished the technical challenges of segmenting the focal plane to allow selecting specific objects for study in a manner that is reconfigurable, dispersing the light and finally detecting it, all while operating at cryogenic temperature at the telescope [\citenum{2003SPIE.4841.1611E}].  FLAMINGOS cryostats were the main innovation.  Slits for segmenting the focal plane were machined and placed in a fore-dewar while the rest of the instrument could be kept cold.  More recent designs use dynamically reconfigurable slit units.  Although the particular approach used in FLAMINGOS was not widely adopted, this instrument served as a proof of concept.  It demonstrated that such a complex system could be made to work at a telescope and it undoubtedly inspired other efforts.  Although this award was acknowledged in the literature only five times, it nevertheless had a profound effect.  Subsequent multi-object spectrographs include the Multi-Object Infrared Camera and Spectrograph (MOIRCS) for the Subaru telescope [\citenum{2008PASJ...60.1347S}],  FLAMINGOS-2 on Gemini [\citenum{2012SPIE.8446E..0IE}], EMIR on Gran Telescopio Canarias [\citenum{2016SPIE.9908E..1JG}], Multi-Object Spectrometer For Infra-Red Exploration (MOSFIRE) on Keck [\citenum{2012SPIE.8446E..0JM}], MMT and Magellan Infrared Spectrograph (MMIRS), which was based explicitly on the FLAMINGOS design [\citenum{2012PASP..124.1318M}], the LBT NIR-Spectroscopic Utility with Camera and Integral-Field Unit for Extragalactic Research (LUCIFER)  on the Large Binocular Telescope [\citenum{2010SPIE.7735E..1LA}], and the K-band Multi-Object Spectrograph (KMOS) on the Very Large Telescope [\citenum{2006SPIE.6269E..1CS,2013Msngr.151...21S}].  

Additional science and technology topics that have been supported by ATI include optical interferometry and optical transients, including the study of gamma-ray burst counterparts and afterglows.  This review has emphasized optical-IR technologies; however, ATI has been equally impactful for submillimeter and radio astronomy.  

\section{CONCLUSION}
\label{ConclusionSection}

ATI has supported transformative research in astronomy over its 30 year history.  It has helped to disseminate key technologies from CCD cameras to more specialized instruments at observatories.  ATI awards have also supported the development of new technologies, including adaptive optics, optical and IR imaging detectors and spectrographs.  

Literature acknowledgement and citation statistics provide a means to assess the impact of ATI awards.  Such analysis shows that ATI awards are acknowledged and the resulting papers are cited in the literature at comparable rates to a similar pure science program that has no technology or instrumentation component.  Thus the direct impact of ATI awards appears comparable to pure science awards. By considering only direct acknowledgements and only peer-reviewed literature, this conclusion may be robust, but it is inevitably conservative.  The total impact of ATI awards has not been captured by literature acknowledgement and citation statistics alone.  A single new instrument may enable dozens of investigations that use the instrument for science; such indirect impacts have not been systematically quantified here.  

Examples provided here illustrate an impact to ATI awards that goes beyond literature citation statistics.  An award may occur at a pivotal time in a larger project, and as a result its impact may only become apparent in hindsight.  Furthermore,  technology maturation may evolve over a longer time period than an individual award.  It may take a decade or more for a technology to mature; but when it does it may open entire frontiers to science.  Finally, opportunities for education and experience provided by ATI awards have not been considered here.  By providing awards that are large enough to make an impact, but small enough for an early career investigator to manage, ATI fills an important niche in the training opportunities for instrument developers.  Taking the long view shows how science benefits from advancements in technology.

\acknowledgments     
 
We acknowledge all previous and current ATI awardees for their efforts in research. For discussions, we thank the following: Christoph Baranec, Jamie Bock, Mark Chun, Scott Diddams, Debra Fischer, Jian Ge, Don Hall, Shaul Hanany, Casey Law, Jared Males, Anthony Readhead, Deqing Ren, Ray Sharples, J. Anthony Tyson.  We also thank Dennis Crabtree for providing data on the median citation rate of publications in {\it The Astronomical Journal}.  


\bibliography{ATIReviewBibliography}   
\bibliographystyle{spiebib}   

\end{document}